# Thermodynamics and spin gap of the Heisenberg ladder calculated by the look-ahead Lanczos algorithm

M. Troyer[a,b,c], H. Tsunetsugu [a,b] and D. Würtz[a,b]

[a]*Interdisziplinäres Projektzentrum für Supercomputing,
Eidgenössische Technische Hochschule, CH-8092 Zürich, Switzerland*
[b]*Theoretische Physik, Eidgenössische Technische Hochschule,
CH-8093 Zürich, Switzerland*
[c]*Centro Svizzero di Calcolo Scientifico, CH-6924 Manno, Switzerland*

We have developed an improved version of the quantum transfer matrix algorithm. The extreme eigenvalues and eigenvectors of the transfer matrix are calculated by the recently developed look-ahead Lanczos algorithm for non-Hermitian matrices with higher efficiency and accuracy than by the power method. We have applied this method to the Heisenberg ladder. The temperature dependence of the susceptibility, specific heat, correlation length and nuclear spin relaxation rate $1/T_1$ are calculated. Our results support the existence of a spin gap of about $0.5J$.

## I. INTRODUCTION

The behavior of one-dimensional (1D) strongly correlated systems and spin chains is by now quite well understood. For two-dimensional (2D) strongly correlated systems there are many open questions. Analytic results are much harder to obtain and there are finite-size scaling problems with numerical methods. Ladder models (double chains) are an interesting intermediate step between 1D and 2D system. They are easier to treat numerically than 2D systems and show new phenomena, which are not present in the 1D chains.[1–6] Another reason for special interest in ladder systems is the possibility of realizing a lattice of weakly coupled ladders in the compounds $Sr_2Cu_4O_6$ and $(VO)_2P_2O_7$.[7,8]

A simple but interesting model is the Heisenberg ladder, consisting of two coupled spin-$\frac{1}{2}$ Heisenberg chains of length $L$:

$$H = J \sum_{a=1,2} \sum_{i=1}^{L} \mathbf{S}_{i,a} \cdot \mathbf{S}_{i+1,a} + J' \sum_{i=1}^{L} \mathbf{S}_{i,1} \cdot \mathbf{S}_{i,2}$$
$$-g\mu_B h \sum_{a=1,2} \sum_{i=1}^{L} S^z_{i,a}. \qquad (1)$$

Here $\mathbf{S}_{i,a}$ is the spin operator at site $i$ ($i = 1, \ldots L$) on the rung $a$ ($a = 1, 2$) and periodic boundary conditions are used along the ladder ($x$-direction). (see Fig. 1). $h$ is an external field in the $z$-direction. The field $h = 0$, except to calculate numerical derivatives with respect to the external field $h$, and we set $g\mu_B = 1$. The exchange constants, $J$ and $J'$, are positive, corresponding to anti-ferromagnetic coupling. As the system is translationally invariant in the $x$-direction the momentum $k_x$ is a good quantum number. In the $y$-direction we use open boundary condition. This is however equivalent to periodic boundary conditions in the $y$-direction and a coupling $J'/2$. The momentum along the rungs $k_y = 0, \pi$ is therefore also well defined.

The Heisenberg ladder shows a completely different behavior than the single chain model. While the excitation spectrum is gapless (des Cloiseaux - Pearson mode[9]) for the spin $S = \frac{1}{2}$ single chain, there exists a spin gap in the ladder,[1–3] similar to the Haldane gap in the $S = 1$ antiferromagnetic Heisenberg chain.[10]

The Heisenberg ladder and related models, such as the $t$-$J$ ladder[2,4] or the Hubbard ladder[5] and most of the interesting strongly correlated quantum systems cannot be solved analytically. Because of strong interactions mean field and perturbation theories often fail to give reliable results either. Actually many interesting phenomena in these models are of non-perturbative origin. Numerical methods giving exact results are thus essential to study such systems.

Four different methods are often used to obtain "exact" numerical results for strongly correlated systems. These are methods without uncontrolled approximations. Two of these methods, quantum Monte Carlo (QMC) and quantum transfer matrix (QTM) work at finite temperatures. The other two, exact diagonalization (ED) for small systems and the density-matrix renormalization-group technique (DMRG) are zero-temperature methods.

The QMC and QTM[11–14] methods are both based on a Trotter-Suzuki decomposition of the partition function.[15] The $d$-dimensional quantum system is mapped onto a $(d+1)$-dimensional classical one. For quasi 1D quantum systems, such as chains or ladders, the partition function can then be obtained by the QTM.[11–14] This method is very powerful. It allows the calculation of the temperature dependence of thermodynamic quantities as well as correlation lengths for *infinite systems*. No extrapolation is thus necessary for the system size. It does not suffer from the "negative sign problem" of QMC and has much higher accuracy.

We have combined the usual QTM method with the look-ahead Lanczos algorithm for non-Hermitian matrices[16] to calculate the the extreme eigenvalues and eigenvectors of the QTM more efficiently. This new method allows us to calculate every thermodynamic quantity with higher accuracy. From the numerical point of view it is much more efficient then the power methods that have usually been used.

In QMC[13] the partition function is calculated by statistical sampling of the corresponding classical system, instead of being evaluated exactly. QMC is very powerful if the "negative sign problem" is not severe. It can be used in any dimension, on systems with more than hundred sites and at lower temperatures than the QTM methods. The results are, however, not as accurate as the QTM results due to statistical errors from the sampling. These errors can be made quite small unless the system investigated suffers from the "negative sign problem". This sign problem, which occurs in many frustrated spin systems and in 2D fermion systems, often makes simulations practically impossible.

Exact diagonalization by the Lanczos algorithm[17] is a very accurate zero-temperature method. It can be used to obtain the ground state and the low lying excitation spectrum for small systems (of up to about $10^8$ states) with high accuracy. However the restriction to small systems often leads to difficulties with finite size scaling. ED can also be used to calculate finite temperature properties. But this requires the calculation of a significant portion of the energy spectrum or of all energy eigenvalues. The QTM method in contrast needs just a few of the extreme eigenvalues of the transfer matrix.

DMRG[18] is another zero-temperature method. It can be used to calculate the ground state and the low lying spectrum for larger systems (about 100 sites). This method works exceptionally well for one-dimensional chains. It can also be applied to higher dimensional systems, but there it is harder to obtain accurate results.

The Heisenberg ladder was studied by Dagotto et al.[2] and Barnes et al.[3] using exact diagonalization of ladders with up to $2 \times 12$ sites and QMC on systems with up to $2 \times 32$ sites. From their finite size results they extrapolated a spin gap of $0.5J$ for the infinite length ladder at the isotropic point $J = J'$. A calculation by Noack et al.[6] using the density matrix renormalization group technique[18] (DMRG) gives a spin gap of $\Delta \approx 0.5037J$ and a correlation length of $\xi = 3.19(1)$ for $J = J'$.

The Heisenberg ladder was also treated in a mean-field approximation by Gopalan et al..[19] They calculate the spin gap and the excitation spectrum. The dispersion of the spin-triplet excitations agrees well with ED results.

Using the QTM method we have studied the temperature dependence of the correlation length $\xi$, the susceptibility $\chi$ and the specific heat $C$ directly for the *infinite* ladder for temperatures down to $T \approx 0.2J$. The spin gap and the temperature dependence of $\xi, \chi, C$ and of the nuclear spin relaxation rate $1/T_1$ at low temperatures was calculated by combining the QTM with ED results on the excitation spectrum.

## II. THE QUANTUM TRANSFER MATRIX METHOD

The QTM method has been widely used to study spin models numerically.[11–13,20] The method is based on a mapping of the $d$-dimensional quantum mechanical system onto a $d+1$ dimensional classical one. For some models, e.g. Bethe ansatz solvable models, the partition function of the corresponding classical model can also be calculated analytically by using a transfer matrix method. The QTM of a 1D spin-1/2 model is equivalent to the diagonal-to-diagonal transfer matrix of the eight-vertex model,[21] which can be solved exactly using Bethe ansatz. 1D models treated analytically include the Heisenberg model,[22] the $XXZ$-model and $XYZ$-model[23,24] and the Hubbard model.[25]

The first step of the QTM is the Trotter-Suzuki decomposition of the grand canonical partition function of a quantum model.[15] The Hamiltonian is decomposed into two parts $H = H_1 + H_2$, each of which is easy to diagonalize. A standard choice is the decomposition into two sums of commuting terms:

$$H_1 = \sum_{i \text{ even}} H^{(i)}, \quad H_2 = \sum_{i \text{ odd}} H^{(i)}, \qquad (2)$$

with $[H^{(i)}, H^{(j)}] = 0$ for $i,j$ both even or both odd. The two sums $H_1$ and $H_2$ do not commute in general. The simplest decomposition for a chain with only nearest neighbor interactions is the so-called "checkerboard decomposition".[26] There all terms on odd-numbered bonds are collected in $H_1$ and the even-numbered bonds into $H_2$ (see Fig. 2a). This is the standard decomposition used in most calculations. We have used it in this paper for the 1D chains. For the 1D Heisenberg model the $H^{(i)}$ are:

$$H^{(i)} = J\mathbf{S}_i \cdot \mathbf{S}_{i+1} - \frac{h}{2}\left(\mathbf{S}_i + \mathbf{S}_{i+1}\right). \qquad (3)$$

A similar decomposition, shown in Fig. 2b, can be used for ladder models. For the Heisenberg ladder it is:

$$\begin{aligned} H^{(i)} = & J \sum_{a=1,2} \mathbf{S}_{i,a} \cdot \mathbf{S}_{i+1,a} \\ & + \frac{J'}{2} \left(\mathbf{S}_{i,1} \cdot \mathbf{S}_{i,2} + \mathbf{S}_{i+1,1} \cdot \mathbf{S}_{i+1,2}\right) \\ & - \frac{h}{2} \sum_{a=1,2} \left(\mathbf{S}_{i,a} + \mathbf{S}_{i+1,a}\right). \end{aligned} \qquad (4)$$

Using this decomposition it is possible to approximate the partition function in the following way:

$$Z = \text{Tr}\left(e^{-\beta H}\right) = \text{Tr}\left((U_1 U_2)^M\right) + \text{O}(\Delta\tau^2)$$
$$= \sum_{i_1,\ldots,i_{2M}} \langle i_1|U_1|i_{2M}\rangle\langle i_{2M}|U_2|i_{2M-1}\rangle \times \cdots$$
$$\times \langle i_3|U_1|i_2\rangle\langle i_2|U_2|i_1\rangle + \text{O}(\Delta\tau^2), \quad (5)$$

where $\beta \equiv 1/T$ denotes the inverse temperature (imaginary time), $M$ is the Trotter number and $\Delta\tau \equiv \frac{\beta}{M}$. The $|i_k\rangle$ are a complete orthonormal system of the states,

$$U_1 = e^{-\Delta\tau H_1} = \prod_{i \text{ even}} e^{-\Delta\tau H^{(i)}},$$
$$U_2 = e^{-\Delta\tau H_2} = \prod_{i \text{ odd}} e^{-\Delta\tau H^{(i)}}. \quad (6)$$

Note that all the factors in each product commute with each other. Since $H_1$ and $H_2$ are chosen to be easy to diagonalize the evaluation of the matrix elements $\langle i|U_1|i'\rangle$ is straightforward.

The decomposition leads to a systematic error which is of order $\Delta\tau^2 \propto M^{-2}$. We can extrapolate to $\Delta\tau \to 0$ ($M \to \infty$) by fitting the results for different Trotter numbers $M$ to a polynomial in $\Delta\tau^2$.

The above equation (5) can be interpreted as an evolution in imaginary time (inverse temperature, also called "Trotter" direction) of the state $|i_1\rangle$ by the "time evolution" operators $U_1$ and $U_2$. Within each time interval $\Delta\tau$ the operators $U_1$ and and $U_2$ are each applied once. This leads to a graphical representation of the sum on a square lattice, where the applications of the operators $U^{(j)} \equiv \exp(-\Delta\tau H^{(j)})$ are marked by shaded squares (see Fig. 3). The configuration on each time slice corresponds to one of the states $|i_k\rangle$ in the sum (5) for $Z$.

The QTM exchanges the space and imaginary time direction. problem is reformulated in terms of column-to-column transfer matrices $V_1$ and $V_2$ as shown in Fig. 3. The partition function can be written similarly to Eq. (5) as

$$Z = \text{Tr}\left[(V_1 V_2)^{L/2}\right] + \text{O}(\Delta\tau^2) = \text{Tr}(V^{L/2}) + \text{O}(\Delta\tau^2), \quad (7)$$

where $V \equiv V_1 V_2$ and $L$ is the length of the chain. Here we have used periodic boundary conditions in the space direction. Again the transfer matrices are products of sparse matrices $V^{(i)}$:

$$V_1 = \prod_{1 < i \leq 2M;\, i \text{ odd}} V^{(i)},$$
$$V_2 = \prod_{1 < i \leq 2M;\, i \text{ even}} V^{(i)}. \quad (8)$$

The matrix $V^{(i)}$ can be calculated quite simply from the corresponding matrices $U^{(j)}$. Let $\sigma_1$ and $\sigma_2$ denote the states on the lower left and right corners of a square and let $\tau_1$ and $\tau_2$ denote the states on the upper corners, as shown in Fig. 4. Then

$$\langle \sigma_2, \tau_2 | V^{(i)} | \sigma_1, \tau_1 \rangle = \langle \sigma_1, \sigma_2 | U^{(j)} | \tau_1, \tau_2 \rangle. \quad (9)$$

In order to describe most of the thermodynamic properties of the system it is enough to know the extreme eigenvalues and eigenvectors of the QTM. This follows from an interchangeability theorem,[14,27] which allows us to interchange the limit of system size $L \to \infty$ and the limit of Trotter number $M \to \infty$. The free energy density (per site or per rung for a single chain or ladder resp.) $f = -\frac{1}{\beta L}\ln Z$ in the thermodynamic limit is:

$$f = -\lim_{L\to\infty}\lim_{M\to\infty} \frac{1}{\beta L} \ln \text{Tr}(V^{L/2})$$
$$= -\frac{1}{2\beta}\lim_{M\to\infty} \ln \Lambda_1,$$

where $\Lambda_1$ denotes the largest eigenvalues of $V$. As we will see later the ratio of the two largest eigenvalues determines the correlation length of the most dominant fluctuation:

$$\xi^{-1} = \lim_{M\to\infty} \frac{1}{2}\ln\left|\frac{\Lambda_1}{\Lambda_2}\right|. \quad (10)$$

All thermodynamic quantities can be calculated as derivatives of the free energy. The magnetic susceptibility could, for example, be calculated as a second derivative of the free energy density $f$ with respect to the magnetic field $h$. However numerically it is much better to calculate it just as a simple derivative of the magnetization. Indeed it is possible to calculate local quantities, such as the magnetization or the internal energy directly from the eigenvectors.

Let us first consider the thermal average of a local quantity such as the $\alpha$-component of the spin $S^\alpha$, the particle density $n$ or the energy density. We will call this observable we want to calculate $A$. We can calculate the thermal average of this quantity anywhere on the lattice due to translational invariance. The effect of the measurement is to locally change one of the weights:

$$\langle A \rangle_L = \frac{1}{Z}\text{Tr}(\mathcal{A}_1 V_2 V^{L/2-1}) = \frac{1}{Z}\text{Tr}(\mathcal{A} V^{L/2-1}), \quad (11)$$

where $\mathcal{A} \equiv \mathcal{A}_1 V_2$. The matrix $\mathcal{A}_1$ is $V_1$ with just the matrix $V^{(1)}$ altered:

$$\mathcal{A}_1 = \mathcal{A}^{(1)} \prod_{3 < i \leq 2M;\, i \text{ odd}} V^{(i)}, \quad (12)$$

where $\mathcal{A}^{(1)}$ is the matrix $V^{(1)}$ modified by the measurement:

$$\langle \sigma_2, \tau_2 | \mathcal{A}^{(1)} | \sigma_1, \tau_1 \rangle = \left\langle \sigma_1, \sigma_2 \left| \frac{AU^{(0)} + U^{(0)}A}{2} \right| \tau_1, \tau_2 \right\rangle. \quad (13)$$

To simplify this further we can rewrite the trace in terms of the right and left eigenvectors $|\psi_i^R\rangle$ resp. $\langle\psi_i^L|$

of the transfer matrix $V$. Let us again exchange the limits $M \to \infty$ and $L \to \infty$. For simplicity we will not write the limit $\lim_{M\to\infty}$ in the following equations, but it is always assumed that this limit is taken. The application of the transfer matrix $V$ projects out the eigenvector of the largest eigenvalue $\Lambda_1$ in the limit $L \to \infty$:

$$\langle A \rangle = \lim_{L\to\infty} \frac{\sum_i \langle \phi_i | V^{L/4} \mathcal{A} V^{L/4-1} | \phi_i \rangle}{\sum_i \langle \phi_i | V^{L/2} | \phi_i \rangle}$$
$$= \frac{\langle \psi_1^L | \mathcal{A} | \psi_1^R \rangle}{\langle \psi_1^L | \psi_1^R \rangle \Lambda_1}. \qquad (14)$$

Thus local quantities are easy to obtain from the eigenvector corresponding to the largest eigenvalue. The specific heat $C$ can now be calculated by a numerical derivative of the internal energy:

$$C = \frac{\partial \langle H \rangle}{\partial T}. \qquad (15)$$

The magnetic susceptibility $\chi$ can be calculated as a numerical derivative of the magnetization with respect to the external field $h$:

$$\chi = \left. \frac{\partial \langle S^z \rangle}{\partial h} \right|_{h=0}. \qquad (16)$$

Similarly we can calculate correlation functions, such as spin correlations. Let us calculate the correlations of the fluctuations of such a quantity around its mean value, $\tilde{A}_i \equiv A_i - \langle A \rangle$, between sites $i$ and $i+d$. In the limit $L \to \infty$ this is:

$$\langle \tilde{A}_i \tilde{A}_{i+d} \rangle = \frac{\langle \psi_1^L | \mathcal{A}_1 (V_2 V_1)^{(d-1)/2} \mathcal{A}_2 | \psi_1^R \rangle}{\langle \psi_1^L | \Lambda_1^{(d+1)/2} | \psi_1^R \rangle}, \qquad (17)$$

for odd $d$, and

$$\langle \tilde{A}_i \tilde{A}_{i+d} \rangle = \frac{\langle \psi_1^L | \mathcal{A}_1 V_2 (V_1 V_2)^{(d-2)/2} \mathcal{M}_1 V_2 | \psi_1^R \rangle}{\langle \psi_1^L | \Lambda_1^{(d+2)/2} | \psi_1^R \rangle}, \qquad (18)$$

for even $d$. These correlations are simple to calculate for short and intermediate ranges $d$. Often more interesting, and much simpler to calculate, is the correlation length, defined as

$$\xi^{-1} = -\lim_{d\to\infty} \frac{1}{d} \ln \langle \tilde{A}_i \tilde{A}_{i+d} \rangle. \qquad (19)$$

As we want to take the limit $d \to \infty$ it is sufficient if we consider the case of even $d$. In the limit $d \to \infty$ formula (18) becomes

$$\lim_{d\to\infty} \langle \tilde{A}_i \tilde{A}_{i+d} \rangle$$
$$= \lim_{d\to\infty} \frac{\langle \psi_1^L | \mathcal{A} | \psi_\alpha^R \rangle \langle \psi_\alpha^L | \mathcal{A} | \psi_1^R \rangle}{\langle \psi_1^L | \psi_1^R \rangle \Lambda_1 \Lambda_\alpha} \left( \frac{\Lambda_\alpha}{\Lambda_1} \right)^{d/2}$$
$$\equiv \lim_{d\to\infty} \frac{\langle \psi_1^L | \mathcal{A} | \psi_\alpha^R \rangle \langle \psi_\alpha^L | \mathcal{A} | \psi_1^R \rangle}{\langle \psi_1^L | \psi_1^R \rangle \Lambda_1 \Lambda_\alpha} \exp\left( -\frac{d}{\xi} + ikd \right). \qquad (20)$$

$\Lambda_\alpha$ is the largest eigenvalue with nonzero overlap $\langle \psi_1^L | \mathcal{A} | \psi_\alpha^R \rangle \langle \psi_\alpha^L | \mathcal{A} | \psi_1^R \rangle$. If the state $\mathcal{A} | \psi_1^R \rangle$ is in the same invariant subspace as $|\psi_1^R\rangle$ (e.g. if $A = S^z$), then it is usually the second largest eigenvalue in this subspace, otherwise (e.g. if $A = S^x$ or $A = S^y$) it is usually the largest eigenvalue in the invariant subspace that contains $\mathcal{M} | \psi_1^R \rangle$. In Eq. (20) it was assumed that there is only one eigenvalue with absolute value $|\Lambda_\alpha|$. The generalization of the above formula to the case of multiple eigenvalues with the same absolute value (e.g. a complex conjugate pair) is straightforward.

The correlation length $\xi$ is

$$\xi^{-1} = \lim_{M\to\infty} \frac{1}{2} \ln \left| \frac{\Lambda_1}{\Lambda_\alpha} \right|. \qquad (21)$$

and the wave vector of the most dominant fluctuation $k$ can be calculated from the phase of $\Lambda_\alpha$:

$$k = \lim_{M\to\infty} \frac{1}{2} \arg \left( \frac{\Lambda_\alpha}{\Lambda_1} \right) + n\pi \qquad (n = 0 \text{ or } 1). \qquad (22)$$

The ambiguity arises because the transfer matrix in this formulation propagates over two sites and cannot distinguish between $k$ and $k + \pi$. It can be resolved by comparing the correlations for odd and even $d$.

## III. THE LOOK-AHEAD LANCZOS ALGORITHM

The numerical problem in the QTM method is the calculation of the extreme eigenvalues and the corresponding eigenvectors of the transfer matrix $V$. This is a similar problem as in exact diagonalization (ED). In ED we want to calculate the lowest eigenvalues and eigenvectors of the Hamiltonian. ED is restricted to small system sizes, as we have to store three vectors of the Hilbert space in the main memory of the computer. In the QTM method we have exchanged the space direction with the imaginary time direction. The length of the chain can now be made as large as one wishes. The price we have to pay is that we have to store the vectors of possible states in the imaginary time direction. We are restricted to a small number of time slices and thus to the high and intermediate temperature regime.

The main problem is that, while both $V_1$ and $V_2$ are hermitian, their product is no longer hermitian, since the two matrices do not commute. Until recently there was no efficient way to calculate eigenvalues and eigenvectors of non-hermitian matrices, since the usual Lanczos algorithm is numerically unstable for non-hermitian matrices, and usually does not converge. Therefore the eigenvalues and eigenvectors were calculated using power methods. Recently however a variant of the Lanczos algorithm, the look-ahead Lanczos algorithm was developed.[16] This is almost always numerically stable and convergent. Very rare exceptions, so-called "incurable breakdowns", can usually be circumvented by using different starting vectors. We have never encountered such an incurable breakdown in our calculations.

The Lanczos algorithm[16,17] is an iterative method to tridiagonalize a matrix $V$. The extreme eigenvalues of the recursively generated tridiagonal matrix converge very rapidly to the eigenvalues of the original matrix. As the matrix $V$ is needed only in form of matrix-vector products $Vv$ the Lanczos algorithm is ideally suited to calculate the extreme eigenvalues and eigenvectors of large, sparse matrices.

The Lanczos algorithm recursively generates the tridiagonal matrix and two sets of vectors $\{v_i\}$ and $\{w_i\}$ ($i = 0 \ldots N - 1$) starting from the vectors $v_0$ and $w_0$. These basis vectors span the $N$-th Krylov subspace of $V$ and $V^\dagger$:

$$\text{span}(\{v_i\}) = \text{span}\{v_0, Vv_0, \ldots, V^{N-1}v_0\}, \quad (23a)$$
$$\text{span}(\{w_i\}) = \text{span}\{w_0, V^\dagger w_0, \ldots, (V^\dagger)^{N-1}w_0\}, \quad (23b)$$

and they are biorthogonal:

$$(v_i, w_j) = \delta_{ij}. \quad (24)$$

The Lanczos algorithm terminates regularly when an invariant subspace of $V$ or $V^\dagger$ has been found and $v_N = 0$ or $w_N = 0$. For non-Hermitian matrices a breakdown occurs, when the the vectors $v_N$ and $w_N$ are orthogonal. Then $v_N \neq 0$ and $w_N \neq 0$, but $(v_N, w_N) = 0$ and the normalization (Eq. (24)) cannot be fulfilled. In finite precision arithmetic there can also be near-breakdowns, when $v_N$ and $w_N$ are nearly orthogonal and the algorithm becomes numerically unstable.

The Lanczos algorithm is most often used for Hermitian matrices, where such breakdowns cannot occur. If we choose $v_0 = w_0$ then we have $v_i = w_i$ for all $i$ since $V = V^\dagger$. The normalization Eq. (24) then becomes simply:

$$(v_N, w_N) = (v_N, v_N) = ||v_N||^2. \quad (25)$$

This is zero only in the case of regular termination, where $v_N = w_N = 0$.

The look-ahead Lanczos algorithm relaxes the condition of tridiagonalizing the matrix. As long as there are no breakdowns or near-breakdowns it is equivalent to the usual Lanczos algorithm. If a breakdown would occur in calculating $v_N$ and $w_N$ it tries to skip over that iteration. The simple three-term recurrence relations of the standard Lanczos algorithm are then replaced by more complex relations including not only the vectors $v_{N-2}, v_{N-1}, Vv_{N-1}$ and $w_{N-2}, w_{N-1}, V^\dagger w_{N-1}$ but also $V^2 v_{N-1}, \ldots, V^l v_{N-1}, (V^\dagger)^2 w_{N-1}, \ldots, (V^\dagger)^l w_{N-1}, \ldots$. $l$ is the look-ahead length. The look-ahead Lanczos algorithm then generates a block-tridiagonal matrix with blocks of size $l$ instead of a tridiagonal one. Usually a look-ahead of $l = 2$ or $3$ is sufficient except in rare cases. In extremely rare cases we would encounter breakdowns with any number of look-ahead steps. This case is called an *incurable breakdown*. For details we refer to the original literature.[16] An implementation of the eigenvalue algorithm is available in electronic form.[28]

The look-ahead Lanczos algorithm allows us to calculate the extreme eigenvalues of the QTM very efficiently and with high accuracy. We need much less iterations compared to the power method. We found that the look-ahead Lanczos algorithm often converges in just a few dozen iterations.

Another advantage is that the eigenvectors can be calculated without any problems by the Lanczos algorithm. This allows us to calculate quantities such as the internal energy or the magnetization directly via Eq. (14). These results are more accurate than the calculation as numerical derivatives of the free energy.

We have compared the algorithm to exact results for the 1-D $XY$ and Heisenberg models.[24] We found that our results are very accurate down to quite low temperatures ($T \approx 0.1J$) for results extrapolated from $M = 1 \ldots 10$.

## IV. RESULTS FOR THE HEISENBERG LADDER

### A. Quantum transfer matrix results

As an application of the new algorithm we have studied the Heisenberg ladder. Specifically we have calculated the correlation length $\xi$, the specific heat $C$ and the magnetic susceptibility $\chi$ as a function of the temperature $T$.

In Fig. 5 we show the susceptibility per spin $\chi$ as a function of the temperature for different values of $J/J'$. At high temperatures the results agree well with a third-order high temperature expansion:

$$\chi(T) = \tfrac{1}{4}T^{-1} - \tfrac{1}{8}(J + \tfrac{1}{2}J')T^{-2} + \tfrac{3}{64}JJ'T^{-3}. \quad (26)$$

At low temperatures we observe an exponential drop of the susceptibility, caused by the gap in the spin excitation spectrum. This drop is steeper for smaller values of $J/J'$, indicating that the gap $\Delta/J'$ decreases with increasing $J$. The spin gap will be studied in more details in Sec. IV B.

The spin gap also leads to an exponential drop of the specific heat, as shown in Fig. 6. At high temperatures there is again good agreement with the high-temperature expansion. The free energy per site is

$$f(T) = -T \ln 2 - \tfrac{3}{32}\left(J^2 + \tfrac{1}{2}JJ'\right)T^{-1}, \quad (27)$$

and the specific heat

$$C(T) = \tfrac{3}{16}\left(J^2 + \tfrac{1}{2}JJ'\right)T^{-2}. \quad (28)$$

In Fig. 7 we show the temperature dependence of the correlation length $\xi$ for the Heisenberg chain and the Heisenberg ladder, calculated by the QTM. The wave vector of the dominant correlation is $\mathbf{k} = (\pi, \pi)$ for the ladder, which corresponds to antiferromagnetic correlations. In the high temperature limit the correlation length is similar in both models. With decreasing temperature the correlation length becomes longer for the ladder. This is because antiferromagnetic correlations are enhanced faster in the ladder due to the larger number of nearest neighbor sites. At low temperatures the correlation length saturates to a finite value, $\xi \approx 3\text{-}4$, which agrees with $\xi \approx 3.19$, determined by the DMRG calculation for zero temperature.[6] This finite correlation length corresponds to a finite spin gap via the relation, $\xi \sim a/\Delta$, where $a$ is a constant of the order of a characteristic spin velocity. In the gapless single chain, on the other hand, the correlation length diverges like $\xi \sim v_s/(\pi T)$ ($v_s$: the spin velocity) as predicted by conformal field theory.[29,30]

### B. Spin gap and low-temperature thermodynamics

To calculate the spin gap and the thermodynamic quantities at low temperatures we start from the limit $J/J' \to 0$, where a simple description of the whole excitation spectrum is available. In that limit, each eigenfunction of the total system can be written as a direct product of one-rung states, which are either spin singlets or one of the triplets ($\sigma = -1, 0, 1$), and the ground state is that with all singlets. Accordingly, each eigenenergy is given by $J'N$, where $N$ is the number of triplet rungs, measured from the ground state energy $-\frac{3}{4}J'L$, and the energy spectrum shows a tower structure consisting of equidistant multiplets with separation $J'$. Each multiplet is labeled by the number of triplet rungs, $N$. The first excited multiplet consists of the states with one triplet rung and therefore belongs to the sector of $S_{\text{tot}} = 1$, and its multiplicity is $3L$. In general, the $N^{\text{th}}$-multiplet, which consists of the states with $N$ triplet rungs, has the multiplicity, $g(L,N) = 3^N \binom{L}{N}$, where the first factor $3^N$ comes from the spin part.

A small but finite value of $J$ lifts the degeneracy of these states. A schematic picture of the energy levels is shown in Fig. 8. The one-triplet excitations then form a three-fold degenerate band of collective excitations with dispersion $\epsilon_k = J' + J \cos k$ and $z$-component of spin $\sigma = -1, 0, 1$, where we set the ground state energy $E_{G.S.} = 0$. The minimum of this band is at a momentum $k_x = \pi$ along the ladder. The momentum along the rung is $k_y = \pi$. We will call these excitations "magnons" although there is no magnetic long range order in the ground state. order . To second order in perturbation theory the gap is[31]

$$\Delta \approx J' - J + \frac{1}{2}J^2/J'. \tag{29}$$

The higher excited states form a continuum of excited states, with $k_y = 0$ and a minimum at $k = 0$. They can be viewed as two-"magnon" states. In low-order perturbation theory the magnon-magnon interaction is repulsive. The minimum of the continuum is thus at energies slightly larger than twice the gap $2\Delta$.

With increasing $J$ the collective one-"magnon" branch crosses into the two-"magnon" continuum (see Fig. 8c). The exact diagonalization[3] and mean-field[19] results indicate that even then the spectrum can still be described by the above picture.

Using these results on the excitation spectrum we can calculate the low temperature thermodynamics of the Heisenberg ladder. First we start from the simple limit $J = 0$. Each rung can be either in the singlet state or in one of the three triplet states. We obtain for the partition function of the ladder of length $L$:

$$Z_0 = (1 + e^{-\beta(\Delta+h)} + e^{-\beta\Delta} + e^{-\beta(\Delta-h)})^L$$
$$= \left[1 + (1 + 2\cosh(\beta h))e^{-\beta\Delta}\right]^L, \tag{30}$$

and for the magnetic susceptibility

$$\chi_0 = \frac{1}{2L\beta}\frac{\partial^2}{\partial h^2}\ln Z\bigg|_{h=0} = \beta\frac{e^{-\beta\Delta}}{1 + 3e^{-\beta\Delta}}, \tag{31}$$

which drops like $\beta e^{-\beta\Delta}$ for low temperatures.

If $J$ is nonzero we have to take into account the dispersion of the spin excitations. We assume the magnon excitations to have a dispersion $\epsilon_k + \sigma h$, where $\sigma = -1, 0, 1$ is the $z$-component of the spin. In the limit $T \to 0$ the interactions between these magnons become negligible since the magnon density goes to zero due to the gap.

The "magnons" are boson-like in the sense that one can excite more than one excitation with the same quantum numbers, the wave number $\mathbf{k}$ and the spin $\sigma$, but they are not real bosons, since the Hilbert space is restricted. One cannot excite two "magnons" at the same rung, which might be described by a hard-core repulsion in the boson representation. This was first pointed out by Dyson for real magnons in a ferromagnetic state, usually referred to as kinematical interactions.[32] The kinematical interactions become important with increasing temperature, and are essential to get a correct temperature dependence. Otherwise, for example, in the limit of $T \to \infty$ the number of bosons at each $\mathbf{k}$ would diverge and we would not obtain the correct entropy for $T \to \infty$.

At low enough temperatures, $T \ll \Delta$, the magnon density is very low and it is sufficient to include up to one magnon for each $\mathbf{k}$ and $\sigma$. Therefore, both residual magnon-magnon interactions and the kinematical interactions are negligible. The free energy per site in that limit is

$$f_1 = -\frac{1}{2L\beta}[1 + 2\cosh(\beta h)]\sum_k e^{-\beta\epsilon_k}$$
$$= -\frac{1}{2\beta}[1 + 2\cosh(\beta h)]z(\beta), \tag{32}$$

where we have replaced the sum by the integral

$$z(\beta) \equiv \frac{1}{2\pi}\int_{-\pi}^{\pi} dk\, e^{-\beta\epsilon_k} = \int_0^\infty d\epsilon\, \rho(\epsilon)e^{-\beta\epsilon}, \tag{33a}$$

which is a Laplace transform of the magnon density of states, $\rho(\epsilon)$. The susceptibility then is

$$\chi_1(T) = -\frac{\partial^2 f_1}{\partial h^2}\bigg|_{h=0} = \beta z(\beta). \quad (34)$$

For a simple form $\epsilon_k = \Delta + a\,||k| - \pi|^n$ we can perform the integration

$$z(\beta) \approx \frac{\Gamma(\frac{1}{n})}{n\pi}(a\beta)^{-1/n} e^{-\beta\Delta}, \quad (35)$$

where we extended the integration over $k$ to infinity. At low temperatures the magnetic susceptibility then is

$$\chi_{(n)}(T) = \frac{\Gamma(\frac{1}{n})}{n\pi} a^{-1/n} T^{-1+1/n} e^{-\Delta/T}. \quad (36)$$

If we replace the magnon band by the quadratic approximation ($n = 2$) we get

$$\chi_{(2)}(T) = \frac{1}{2\sqrt{\pi a T}} e^{-\Delta/T}. \quad (37)$$

Similarly we can calculate the specific heat as

$$C_{(n)}(T) = \frac{3}{2n\pi}\left(\frac{\Delta}{a}\right)^{1/n}\left(\frac{T}{\Delta}\right)^{2-1/n} e^{-\Delta/T}$$
$$\times \left[\Gamma(\tfrac{1}{n}) + 2\Gamma(1+\tfrac{1}{n})\frac{T}{\Delta} + \Gamma(2+\tfrac{1}{n})\left(\frac{T}{\Delta}\right)^2\right]. \quad (38)$$

In the quadratic approximation:

$$C_{(2)}(T) = \frac{3}{4}\left(\frac{\Delta}{\pi a}\right)^{1/2}\left(\frac{T}{\Delta}\right)^{3/2}$$
$$\times \left[1 + \frac{T}{\Delta} + \frac{3}{4}\left(\frac{T}{\Delta}\right)^2\right] e^{-\Delta/T}. \quad (39)$$

The low-temperature result Eq. (36) motivates a first estimate of the gap based on the logarithmic derivative $-\frac{\partial \ln \chi}{\partial \beta}$. This derivative is $\Delta - (1 - 1/n)T$ at low temperatures for a susceptibility (36). This derivative goes to zero on the other hand if the susceptibility is nonzero for $T = 0$ or vanishes following a power law.

In Fig. 9 we show $-\frac{\partial \ln \chi}{\partial \beta}$ for some gapless systems, the 1D Heisenberg and XY-models, to compare with the Heisenberg ladder. This plot clearly shows the existence of a spin gap for the ladder.

The size of the gap however is not easy to estimate from the data. For $J = J'$ we can reach only temperatures $T/J' \approx 0.2$, which is below the gap but not yet in the asymptotic region.

To determine the size of the spin gap more precisely, we need a fitting function which describes the whole temperature range. This function should give correct results in both low and high temperature limits. To get a correct high temperature limit, one has to take into account the kinematic interactions, as discussed before. We have found a simple way of including kinematical interactions in the thermodynamics based on reasonable physical arguments. Our formula not only gives correct low and high temperature limits, but the overall agreement also turns out to be nice.

In our new formula, the grand partition function is calculated as follows. The main problem of the boson description is that as the number of triplet rungs, $N$, increases, the number of the corresponding boson basis states diverges like $g_B(L, N) = \binom{3L+N-1}{N}$, while the correct dimension is $g(L, N) = 3^N \binom{L}{N}$. Therefore, the basic idea is to reweight the $N$-magnon part in the partition sum, $[g(L, N)/g_B(L, N)]Z_{\text{boson}}(N\text{-magnon})$, so that each multiplet contributes the correct entropy.[33] This can be done with slight modification of the boson partition function,

$$Z'_C(L, N) = \sum_{\{k_j, \sigma_j\}} \exp\left[-\beta \sum_{j=1}^{N}(\epsilon_{k_j} - h\sigma_j)\right]. \quad (40)$$

This corresponds to a sum neglecting the undistinguishability of bosons. Aside from the global factor, the difference from the original boson sum are the terms in which two or more bosons have the same quantum numbers, $(\mathbf{k}, \sigma)$. However, the number of these terms is smaller by at least order one w. r. t. $L$ and we neglect those corrections. Since there are $(3L)^N$ terms in $Z'_C$, the reweighting should work as follows:

$$\tilde{Z} \equiv \sum_{N=0}^{L} \frac{g(L, N)}{(3L)^N} Z'_C(L, N)$$
$$= \sum_{N=0}^{L} \binom{L}{N} L^{-N} \sum_{k_1,\ldots,k_N}(1 + 2\cosh(\beta h))^N e^{-\beta \sum_{i=1}^{N} \epsilon_{k_i}}$$
$$= \left[1 + (1 + 2\cosh(\beta h))\frac{1}{L}\sum_k e^{-\beta \epsilon_k}\right]^L. \quad (41)$$

There are $4^L$ terms in total in the above partition sum, giving the correct total entropy, since we reweighted to get the correct number of excitations. Note that here (i) we assume that all excitations could be described as multi-magnon excitations and (ii) all residual magnon-magnon interactions are neglected. The assumption (i) is obviously correct in the limit of $J/J' \to 0$ and there is no indication of a breakdown of the arguments of analytic continuation w. r. t. $J$: e.g., the spin gap is always finite as far as $J$ is nonzero.

The free energy per site is

$$\tilde{f} = -\frac{1}{2\beta}\ln\left[1 + (1 + 2\cosh(\beta h))z(\beta)\right], \quad (42)$$

where we have taken the limit $L \to \infty$ and again replaced the sum over $k$ by an integral.

This partition functions gives a susceptibility

$$\tilde{\chi} = \beta \frac{z(\beta)}{1 + 3z(\beta)}, \qquad (43)$$

which is correct in both limits $T \to 0$ and $T \to \infty$. For very low temperatures we recover the result of the low-temperature approximation (34). For high temperatures we obtain the correct Curie law $\tilde{\chi} \approx \frac{1}{4T}$.

We will now try to fit the QTM results to the above model. The function $z(\beta)$ depends on the dispersion $\epsilon_k$ we use. First we discuss the small $J/J'$ region. As the correction term in the Trotter-Suzuki decomposition is of order $\beta^3 J^3/M^2$ we can reach quite low temperatures $T/J'$ when $J \ll J'$. For $J = 0.1 J'$ we can reach temperatures below $T/J' \approx 0.04$. These temperatures are low enough to see the asymptotic behavior. In that limit the dispersion is of the form

$$\epsilon_k = J' + J \cos k = \Delta + 2J \cos^2(k/2), \qquad (44)$$

with $\Delta = J - J'$. For $J/J' = 0.1$ we can reach quite low temperatures and a fit to the above Eq. (43) using the dispersion (44) is excellent. The resulting fit of $\chi(T)$ is shown in Fig. 10. A least square fit gives a gap of $\Delta = 0.909$, which is in excellent agreement with the second order perturbation result $\Delta = 0.905$ (Eq. (29)).

At $J = J'$ the gap is harder to estimate. This is caused by two facts. First we cannot reach as low temperatures as in the small $J$ region, as the Trotter "time" step $\beta J/M$ is now larger. The lowest temperatures we can reach are about $T/J \approx 0.2$. Additionally we do not know the exact shape of the dispersion.

We can guess the form of the dispersion from exact diagonalization data[3] and mean-field calculations.[19] The dispersions obtained by both the perturbation result Eq. (44) and the mean-field are quadratic close to the minimum at $k = \pi$. At larger $|k - \pi|$ exact diagonalization and mean-field results indicate a more linear behavior. We have used several functional forms for the dispersion. Good fits were obtained by the following dispersions:

$$\epsilon_k^{(1)} = \sqrt{\Delta^2 + 4\Delta a(1 + \cos k)^2}, \qquad (45a)$$

$$\epsilon_k^{(2)} = \sqrt{\Delta^2 + 2\Delta a(k - \pi)^2}, \qquad (45b)$$

$$\epsilon_k^{(3)} = \begin{cases} \Delta + a(|k| - \pi)^2 & \text{if } ||k| - \pi| < \frac{c}{2a}, \\ \Delta - \frac{c^2}{2a} + c||k| - \pi| & \text{otherwise,} \end{cases} \qquad (45c)$$

$$\epsilon_k^{(4)} = \Delta + c||k| - \pi|. \qquad (45d)$$

The dispersion (45a) is the functional form obtained by the mean-field calculation.[19]

In Table I we show the gap, the curvature $a = \frac{1}{2}\frac{\partial^2 \epsilon_k^{(i)}}{\partial k^2}\big|_{k=\pi}$ and the other fitting parameters obtained by a least square fit of the QTM results for $\ln \chi$ to the above dispersions (45).

The discrepancies between the fits arise because, although we can simulate at temperatures below the gap $T \approx 0.4\Delta$, we are not really in the low-temperature regime where the interactions between the excitations and the exact shape of the dispersion become unimportant. This can be seen best in the plot of $-\frac{\partial \ln \chi}{\partial \beta}$ in Fig. 9. In Fig. 11 we show the fit of the susceptibility for the dispersion (45a). The susceptibilities obtained using the other dispersions differ only slightly.

The dispersion $\epsilon_k^{(4)}$ is not realistic, as it is not quadratic, but linear close to the minimum at $k = \pi$. It underestimates the gap, since the density of states is too small near the minimum. For the same reason we believe that the dispersion $\epsilon_k^{(3)}$ underestimates the correct gap. Similarly a dispersion that is too flat close to the minimum overestimates the density of states there and thus also the gap. We estimate the gap to be in the range $0.45J < \Delta < 0.5J$, which is in agreement with the exact diagonalization[3] and DMRG results.[6]

### C. Nuclear spin relaxation rate

Another quantity of interest is the nuclear spin relaxation rate $1/T_1$. It can be written in terms of the dynamical susceptibility perpendicular to the field:[34]

$$\frac{1}{T_1} = 2\gamma^2 T \sum_{\mathbf{q}} |A_{\mathbf{q}}|^2 \frac{\chi''_\perp(\mathbf{q}, \omega_0)}{\omega_0}, \qquad (46)$$

where $|A_{\mathbf{q}}|^2 \propto (2L)^{-1}$ is the form factor, $\gamma$ is the nuclear gyro-magnetic ratio, and $\omega_0$ the nuclear resonance frequency, which is a very small energy scale, typically of the order of mK. The main problem here is the calculation of the imaginary part of the susceptibility $\chi''_\perp(\mathbf{q}, \omega_0)$. This can be related to the dynamical structure factor by the fluctuation-dissipation theorem:

$$\chi''_\perp(\mathbf{q}, \omega_0) = \mathcal{S}_\perp(\mathbf{q}, \omega_0)(1 - e^{-\beta \omega_0}) \approx \mathcal{S}_\perp(\mathbf{q}, \omega_0) \beta \omega_0, \quad (47)$$

where we have used the fact that $\omega_0 \approx 3\text{mK} \ll T$. As the Hamiltonian of the system is invariant under spin rotations and there is no long range order present the susceptibilities in all directions are equal: $\chi_\perp = \chi_{zz}$ and

$$\mathcal{S}_\perp(\mathbf{q}, \omega_0) = \mathcal{S}_z(\mathbf{q}, \omega_0) \qquad (48)$$
$$= \sum_{m,n} |\langle m|S_{\mathbf{q}}^z|n\rangle|^2 \delta(E_m - E_n - \omega_0) e^{-\beta E_n}/Z,$$

where $|m\rangle, |n\rangle$ are complete sets of eigenstates with energy $E_m$ and $E_n$ respectively, and

$$S_{\mathbf{q}}^z \equiv \frac{1}{\sqrt{2L}} \sum_{\mathbf{r}} e^{i\mathbf{q} \cdot \mathbf{r}} S_{\mathbf{r}}^z. \qquad (49)$$

Which states contribute to $1/T_1$ at low temperatures? The dominant fluctuations in the ground state are antiferromagnetic, leading to a maximum in the equal-time spin structure factor at $\mathbf{q} = (\pi, \pi)$. However these dominant antiferromagnetic fluctuations $\langle m|S_{\mathbf{q}}^z|G.S.\rangle$ of the ground state near $\mathbf{q} = (\pi, \pi)$ do not contribute since they have an energy gap of $E_m - E_{G.S.} > \Delta \gg \omega_0$. The only relevant contributions arise from fluctuations of the excited states with small momentum transfer $\mathbf{q}$.

At low temperatures we assume the one-magnon states to be independent. We restrict the sum over $n$ to the independent one-magnon states $|k, \sigma\rangle$ with momentum $k$ and $z$-component of spin $\sigma = -1, 0, 1$. As $\omega_0 \ll \Delta$ and momentum is conserved only the states $|k + q_x, \sigma\rangle$, contribute to the sum over $m$:

$$\mathcal{S}_z(\mathbf{q}, \omega_0) \approx \sum_{k,\sigma} |\langle k + q_x, \sigma | S_{\mathbf{q}}^z | k, \sigma \rangle|^2 \delta_{q_y, 0}$$
$$\times \delta(\epsilon_{k+q_x} - \epsilon_k - \omega_0) e^{-\beta \epsilon_k}, \qquad (50)$$

where $|k, \sigma\rangle$ is a one-magnon state with momentum $k$ and $z$-component of spin $\sigma = -1, 0, 1$. From the excitation spectrum it is obvious that only for $q_y = 0$ and $q_x \approx 0$ or $q_x \approx -2k$ we have a nonvanishing $\mathcal{S}_z(\mathbf{q}, \omega_0)$. Using a second order Taylor expansion of the dispersion we can write $\mathcal{S}_z(\mathbf{q}, \omega_0)$ in terms of $\delta$-functions of $\mathbf{q}$:

$$\mathcal{S}_z(\mathbf{q}, \omega_0) \approx \sum_{k,\sigma} |\langle k + q_x, \sigma | S_{\mathbf{q}}^z | k, \sigma \rangle|^2 \delta_{q_y, 0} e^{-\beta \epsilon_k}$$
$$\times \frac{\delta(q_x) + \delta(q_x + 2k)}{|v(k)| \sqrt{1 - 2\omega_0 \frac{\partial}{\partial k} \frac{1}{v(k)}}}, \qquad (51)$$

where we have set $\omega_0 \to 0$ in the $\delta$-functions. $v(k) = \frac{\partial \epsilon_k}{\partial k}$ is the group velocity. The matrix elements are:

$$|\langle k, \sigma | S_{(0,0)}^z | k, \sigma \rangle|^2 = \frac{1}{2L} \sigma^2. \qquad (52)$$

We estimated the matrix element $|\langle -k, \sigma | S_{-2k, 0}^z | k, \sigma \rangle|^2$ by exact diagonalization on finite ladders of up to 10 rungs. For $J = J'$ is is nearly constant for $\pi/2 < |k| < \pi$:

$$|\langle -k, \sigma | S_{(-2k, 0)}^z | k, \sigma \rangle|^2 \approx 0.5 \frac{1}{2L} \sigma^2. \qquad (53)$$

At low temperatures the main contributions arise from $k \approx \pi$, where $\mathbf{q} \approx (0, 0)$. We replace the matrix elements $|A_q|^2$ by its value $|A_q|^2 \approx A_0^2 \equiv A^2/2L$ at $\mathbf{q} = (0, 0)$. Replacing sums by integrals we get in the quadratic approximation for the dispersion in the temperature range $\omega_0 \ll T \ll \Delta$:

$$\frac{1}{T_1} = \frac{2\gamma^2 A^2}{4\pi} \sum_{q_y = 0, \pi} \int dq_x \mathcal{S}_\perp(\mathbf{q}, \omega_0) \qquad (54)$$

$$\approx \frac{\gamma^2 A_0^2}{8\pi^2} \int dk \frac{3 e^{-\epsilon_k/T}}{2a \sqrt{(\pi - k)^2 + \omega_0/a}} \qquad (55)$$

$$\approx \frac{3\gamma^2 A^2}{16 a \pi^2} e^{-\Delta/T} K_0 \left( \frac{\omega_0}{2T} \right), \qquad (56)$$

where $K_0$ is the modified Bessel function of second kind. In the temperature regime where our approximation is valid we can expand $K_0(\frac{\omega_0}{2T}) \approx -\mathbf{C} + \ln 4 - \ln(\omega_0/T) \approx$ 0.80908 $- \ln(\omega_0/T)$. $\mathbf{C} \approx 0.577216$ is Euler's constant. Thus finally we have for the nuclear spin relaxation rate

$$\frac{1}{T_1} \approx \frac{3\gamma^2 A^2}{16 a \pi^2} e^{-\Delta/T} (0.80908 - \ln(\omega_0/T)), \qquad (57)$$

in the temperature range $\omega_0 \approx 3\text{mK} \ll T \ll \Delta$. The main feature is the exponential drop with temperature caused by the gap. In addition there is a logarithmic divergence in $\omega_0$ caused by the van Hove singularity at the band minimum in the density of states of spin excitations. Although the equal time spin correlations have a maximum at $\mathbf{q} = (\pi, \pi)$ these fluctuations do not contribute since they have a large energy gap. As the main contribution to the nuclear spin relaxation rate comes from $\mathbf{q} \approx (0, 0)$, and not from $\mathbf{q} \approx (\pi, \pi)$ we expect the temperature dependence to be similar for Cu and O sites in a copper-oxide ladder. This differs from the case of copper-oxide planes, where there is a marked difference in the temperature dependence, because there are low energy fluctuations around $\mathbf{q} = (\pi, \pi)$ that contribute to $1/T_1$ at Cu sites but not at O sites.[35]

## V. CONCLUSIONS

We have developed an improved version of the quantum transfer matrix algorithm. Quantum transfer matrix methods (QTM) do not suffer from the sign problem of quantum Monte Carlo. Therefore they are ideal to investigate models where the sign problem is severe. Examples include frustrated spin systems or fermionic ladder models, like the $t$-$J$ ladder.

We have combined the QTM method with the look-ahead Lanczos algorithm to calculate the extreme eigenvalues and eigenvectors of the QTM with very high accuracy. The algorithm converges much faster than usual power methods. The calculation of the *eigenvectors* of the transfer matrix with high precision by the look-ahead Lanczos algorithm allows a direct calculation of the magnetization, internal energy, magnetic susceptibility, specific heat and similar quantities for an *infinite* length system.

In this paper we have reported on the thermodynamics of the Heisenberg ladder. The QTM method by itself is restricted to high and intermediate temperatures ($T > 0.2J$). By combining the QTM method with exact diagonalization results for the low-lying excitation spectrum we are able to calculate the temperature dependence of the specific heat, magnetic susceptibility and the correlation length for the entire temperature range. This also allows an estimation of the spin gap. Finally we have calculated the temperature and frequency dependence of the nuclear spin relaxation rate $1/T_1$.

In the final stages of the preparation of the manuscript we learned about a preprint by Barnes and Riera, in which they report on a calculation of the temperature dependence of the magnetic susceptibility by exact diagonalization.[36]

An interesting question arising here is, what happens to the spin gap upon doping of holes ($t$-$J$-ladder model). This is currently being investigated[2,4,5]

## ACKNOWLEDGMENTS

This work was supported by the NFP-30 program of the Swiss National Science Foundation and by an internal grant of ETH Zürich. The authors wish to thank M.H. Gutknecht, M. Imada, H. Monien, and especially T.M. Rice for helpful discussions. The calculations were performed on the Cray Y-MP/464 of ETH Zürich and on the NEC SX-3/24R of the Centro Svizzero di Calcolo Scientifico CSCS Manno.

[1] R. Hirsch, Diplomarbeit Universität Köln, 1988.
[2] E. Dagotto, J. Riera, and D. J. Scalapino, Phys. Rev. B **45** 5744 (1992).
[3] T. Barnes, E. Dagotto, J. Riera and E.S. Swanson, Phys. Rev. B **47** 3196 (1993).
[4] H. Tsunetsugu, M. Troyer and T.M. Rice, Phys. Rev. B, in press.
[5] R.M. Noack, S.R. White and D.J. Scalapino, "Correlations in a two-chain Hubbard model", preprint cond-mat/9401013.
[6] S.R. White and R.M. Noack "Resonating Valence Bond Theory of Coupled Heisenberg Chains", preprint cond-mat/9403042
[7] M. Takano et al., JJAP Series **7**, 3 (1992).
[8] D. C. Johnston et al., Phys. Rev. B **35**, 219 (1987).
[9] J. des Cloiseaux and J.J. Pearson, Phys. Rev. **128**, 2131 (1962).
[10] For a review, see e.g., I. Affleck, J. Phys.: Condens. Matter **1**, 3047-3072 (1989).
[11] I. Morgenstern and D. Würtz, Phys. Rev. B **32**, 532 (1985); H. Betsuyaku, Prog. of Theor. Phys. **75**, 774 (1986).
[12] H. Betsuyaku, Prog. Theor. Phys. **73**, 319 (1985); T. Yokota and H. Betsuyaku, Prog. Theor. Phys. **75**, 46 (1986).
[13] For an overview see papers in *Quantum Monte Carlo Methods*, edited by M. Suzuki, (Springer Verlag, Berlin, 1987), and references given therein.
[14] M. Suzuki, Phys. Rev. B **31**, 2957 (1985).
[15] H.F. Trotter, Proc. Am. Math. Soc. **10**, 545 (1959); M. Suzuki, Prog. of Theor. Phys. **56**, 1454 (1976).
[16] M. H. Gutknecht, SIAM J. Matrix Anal. Appl. **13**, 594 (1992); ibid. (Jan. 1994); Roland W. Freund, Martin H. Gutknecht and N. Nachtigal, SIAM J. Sci. Comput., **14** 137 (1993).
[17] See, for example, Jane K. Cullum and Ralph A. Willoughby, *Lanczos Algorithms for Large Symmetric Eigenvalue Computations*, (Birkhäuser Verlag, Boston, 1985).
[18] S.R. White, Phys. Rev. Lett. **69**, 2863 (1992); Phys. Rev. B **48**, 10345 (1993).
[19] S. Gopalan, T.M. Rice and M. Sigrist, Phys. Rev. B **49**, 8901 (1994).
[20] See e.g. S. Takada and K. Kubo, J. Phys. Soc. Jpn. **55**, 1671 (1986); K. Kubo and S. Takada, ibid. **55**, 438 (1986); H. Betsuyaku and T. Yokota, Prog. Theor. Phys. **75**, 808 (1986); T. Delica, K. Kopinga, H. Leschke, K.K. Mon, Europhys. Lett. **15**, 55 (1991); K. Kubo, Phys. Rev. B **46**, 866 (1992).
[21] R.J. Baxter, *Exactly solved models in statistical mechanics* (Academic press 1982).
[22] T. Koma, Prog. Theor. Phys. **78**, 1213 (1987); M. Yamada, J. Phys. Soc. Jpn. **59**, 848 (1990).
[23] T. Koma, Prog. Theor. Phys. **81**, 783 (1989).
[24] M. Takahashi, Phys. Rev. B **43**, 5788 (1991); Erratum, Phys. Rev. B **44**, 5397 (1991); M. Takahashi, *ibid* B **44**, 12382 (1991).
[25] R.Z. Bariev, Teor. Mat. Fiz. **49**, 261 (1981) [Theo. Math. Phys. **49**, 1021 (1982)]; T. Koma, Prog. Theor. Phys. **83**, 655 (1990); H. Tsunetsugu, J. Phys. Soc. Jpn. **60**, 1460 (1991).
[26] M. Barma and B.S. Shastry, Phys. Lett. **61** A, 15 (1977); Phys. Rev. B **18**, 3351 (1978).
[27] M. Suzuki and M. Inoue, Prog. Theor. Phys. **78**, 787 (1987).
[28] The source code of the look-ahead Lanczos algorithm for the eigenvalue computation is available as part of the "linalg/lalqmr" package from netlib. It can be accessed by anonymous ftp to "netlib.att.com".
[29] As a review, J. L. Cardy, in *"Phase Transitions and Critical Phenomena"*, vol. 11, (eds.) C. Domb and J. L. Lebowitz, (Academic, London, 1987).
[30] K. Nomura and M. Yamada, Phys. Rev. B **43**, 8217 (1991).
[31] M. Reigrotzki, Diplomarbeit ETH, 1994. In the paper by Barnes et al.[3] a term was omitted in the strong coupling expansion. They obtained $\Delta = J' - J + \frac{3}{4}\frac{J^2}{J'}$.
[32] F.J. Dyson, Phys. Rev. **102**, 1217 and 1230 (1956).
[33] An alternative way of restricting boson number is to introduce a temperature dependent chemical potential. This was first done for a 1D ferromagnet by using the modified spin wave approximation, [M. Takahashi, Phys. Rev. Lett. **58**, 168 (1987)], and then for antiferromagnets by the same method, [M. Takahashi, Phys. Rev. B **40**, 2494 (1989)], and by the Schwinger boson method, [D.P. Arovas and A. Auerbach, Phys. Rev. B **38**, 316 (1988)].
[34] T. Moriya, J. Phys. Soc. Jpn. **18**, 516 (1963).
[35] See e.g. M. Takigawa *et al.*, Phys. Rev. B **43**, 247 (1991).
[36] T. Barnes and J. Riera, "The susceptibility and excitation spectrum of $(VO)_2P_2O_7$ in ladder and dimer chain models", preprint cond-mat/9404060.

TABLE I. Gap and fitting parameters obtained by fitting the numerical data for the susceptibility using various dispersions.

| dispersion | $\Delta/J$ | $a/J$ | $c/J$ |
|---|---|---|---|
| $\epsilon_k^{(1)}$ | 0.496 | 3.17 | — |
| $\epsilon_k^{(2)}$ | 0.517 | 2.62 | — |
| $\epsilon_k^{(3)}$ | 0.438 | 8.21 | 1.30 |
| $\epsilon_k^{(4)}$ | 0.395 | — | 1.29 |

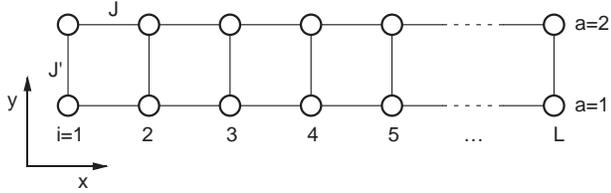

FIG. 1. Diagram of the Heisenberg ladder with two legs in the $x$ direction and $L$ rungs in the $y$ direction. The coupling along the legs is $J$ and the along the rungs $J'$.

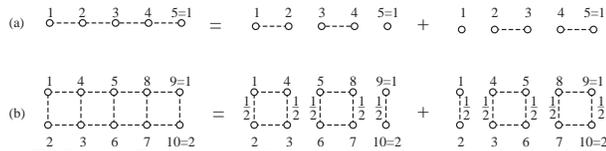

FIG. 2. Examples of decompositions used in the Trotter-Suzuki decomposition. (a) the checkerboard decomposition, the simplest decomposition for 1D chains. (b) a "checkerboard" decomposition for ladder models.

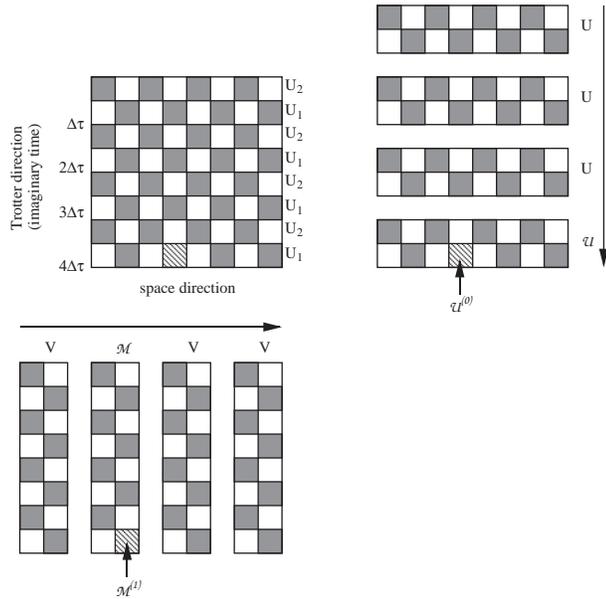

FIG. 3. Graphical representation of the Trotter-Suzuki decomposition of a one dimensional quantum chain using the checkerboard decomposition. Also shown is the formulation in terms of the usual row to row transfer matrices $U$ and in terms of column to column transfer matrices $V$. The matrices that are altered for measurements are indicated by a lighter shading and are labeled. Refer to the text for details.

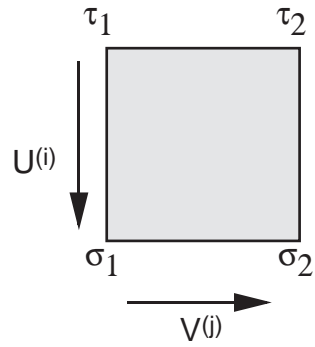

FIG. 4. Rotation of the transfer matrix: The matrix $U^{(i)}$ propagates a state along the imaginary time direction. $V^{(j)}$ propagates along the space direction.

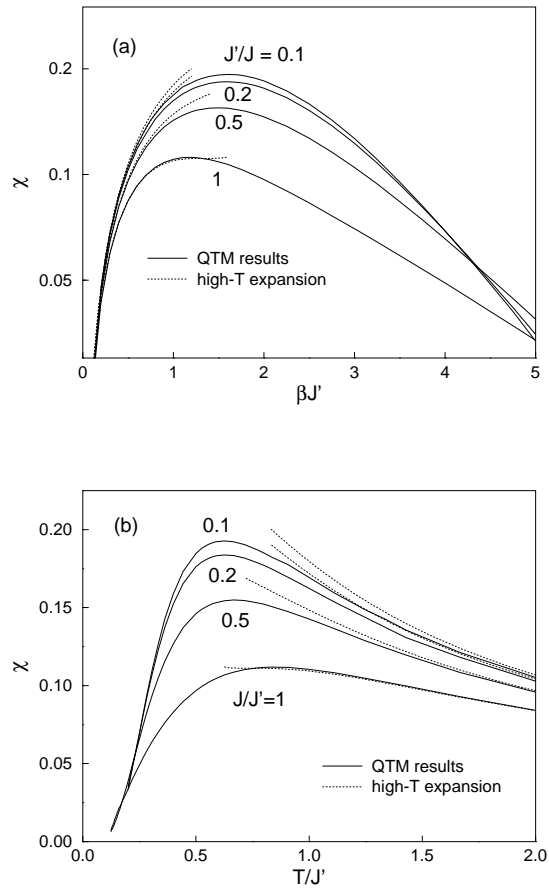

FIG. 5. Temperature dependence of the magnetic susceptibility of the Heisenberg ladder for different values of $J/J' = 1, 0.5, 0.2$ and $0.1$. (a) a logarithmic plot of $\chi$ as a function of the inverse temperature $\beta$. (b) $\chi$ as a function of the temperature $T$.

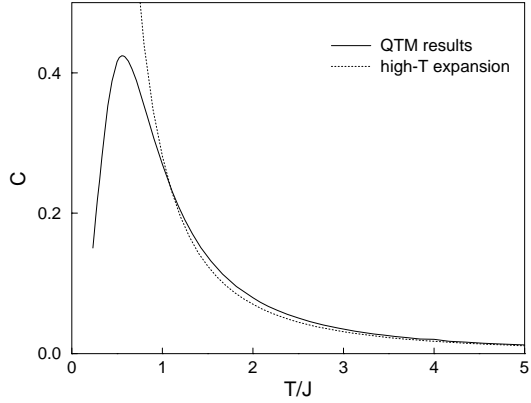

FIG. 6. Temperature dependence of the specific heat of the Heisenberg ladder for $J' = J$.

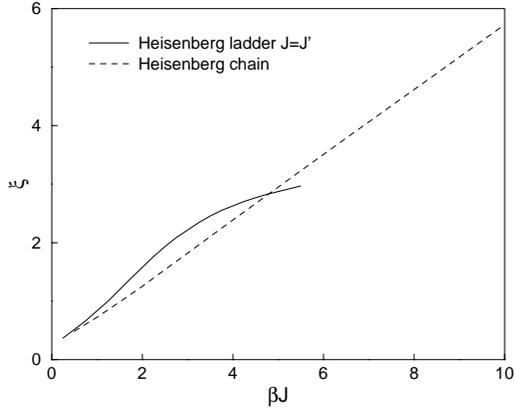

FIG. 7. Correlation length of the Heisenberg chain and ladder as a function of temperature for $J' = J$. In the gapless Heisenberg chain $\xi$ diverges for $T \to \infty$, while it remains finite for the Heisenberg ladder which exhibits a spin gap.

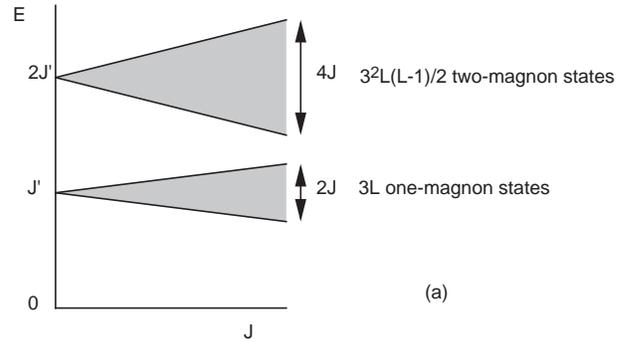

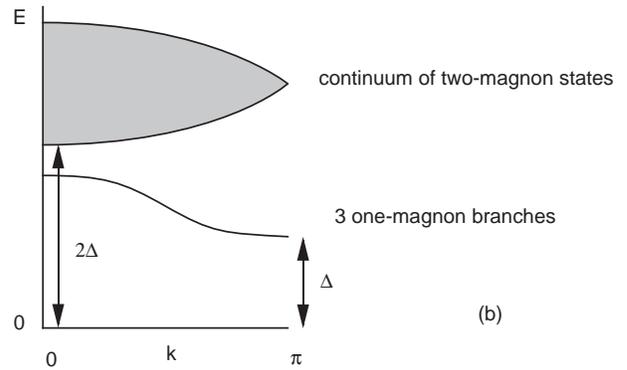

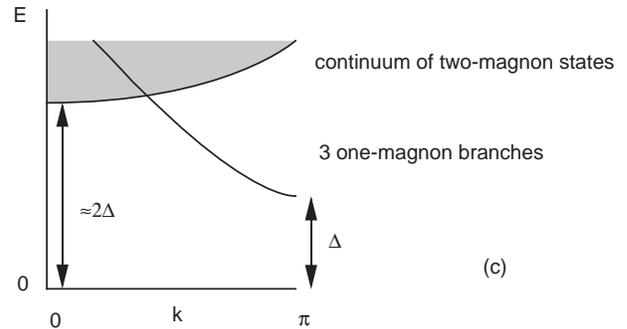

FIG. 8. (a) Evolution of the energy levels when the interaction $J$ is turned on. (b) Qualitative picture of the dispersion at small $J/J'$; (c) at $J = J'$.

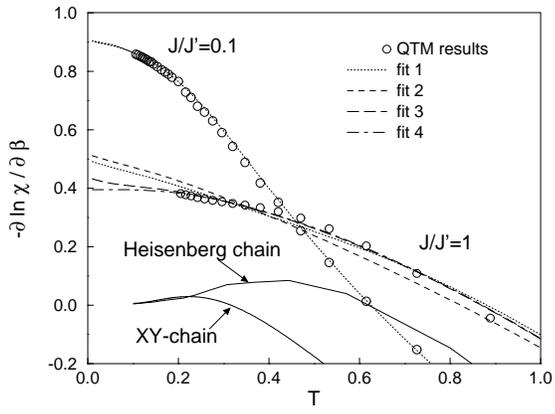

FIG. 9. Temperature dependence of the logarithmic derivative of the magnetic susceptibility with respect to the inverse temperature $\beta$. Shown is the magnetic susceptibility for the gapless 1D $XY$ and Heisenberg chain and for the gapful Heisenberg ladder with $J/J' = 1$ and $J/J' = 0.1$. Also included is the fit which is described in the text. For $J/J' = 1$ the four different fits are shown. The temperature is in units of $J$ for the single chains, and in units of $J'$ for the ladder.

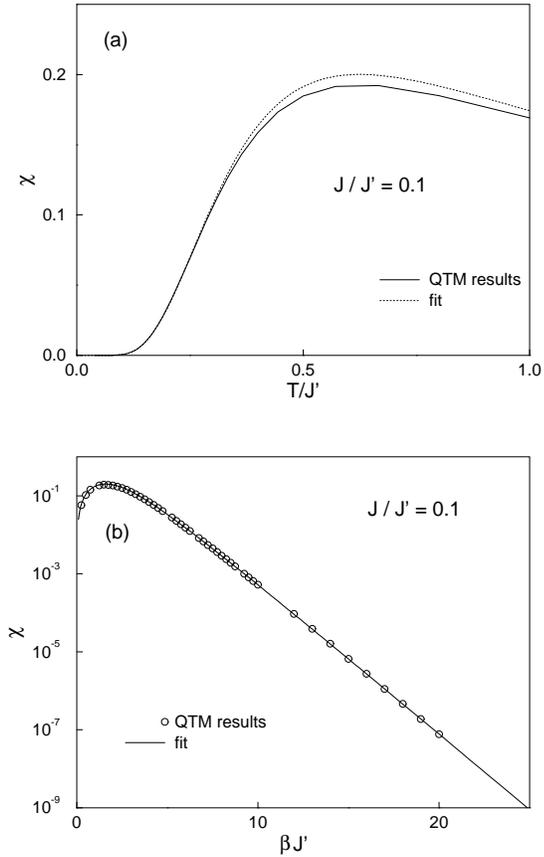

FIG. 10. Temperature dependence of the magnetic susceptibility of the Heisenberg ladder with $J/J' = 0.1$. (a) $\chi$ as a function of the temperature $T$; (b) a logarithmic plot of $\chi$ as a function of the inverse temperature $\beta$.

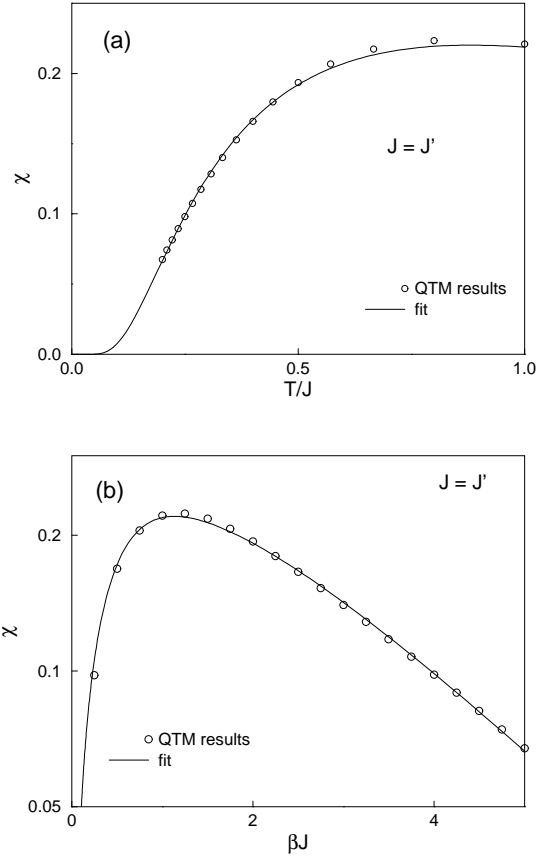

FIG. 11. Temperature dependence of the magnetic susceptibility of the Heisenberg ladder with $J' = J$. The solid line is the fit discussed in the text. (a) $\chi$ as a function of the temperature $T$; (b) a logarithmic plot of $\chi$ as a function of the inverse temperature $\beta$.